\documentclass[12pt]{article}

\usepackage{scicite}
\usepackage{graphicx}
\usepackage{times}
\usepackage{bm}
\usepackage{lipsum}
\newcommand*\VF[1]{\mathbf{#1}}
\usepackage{amsmath,bm}
\usepackage{xcolor}
\usepackage{siunitx}
\usepackage{amsmath}
\usepackage{multirow}
\usepackage{bigints}
\newcommand*\dif{\mathop{}\!\mathrm{d}}
\usepackage{xcolor}

\newenvironment{sciabstract}{%
\begin{quote} \bf}
{\end{quote}}

\title{Experimental Realization of Convolution Processing in Photonic Synthetic Frequency Dimensions}

\author
{Lingling Fan,$^{\dagger}$ Kai Wang,$^{\dagger, \ddagger}$ Heming Wang,$^{\dagger}$ Avik Dutt,$^\mathsection
$ Shanhui Fan$^{\dagger\ast}$\\
\normalsize{$^{\dagger
}$Department of Electrical Engineering, Ginzton Laboratory, Stanford University,}\\ \normalsize{Stanford, CA 94305, USA}\\
\normalsize{$^{\ddagger}$Department of Physics, McGill University,  3600 Rue University, Montreal,}\\ \normalsize{Quebec H3A 2T8, Canada}\\
\normalsize{$^{\mathsection}$Department of Mechanical Engineering and Institute for Physical Science and Technology,}\\ \normalsize{University of Maryland, College Park, Maryland 20742, USA}\\
\\
\normalsize{$^\ast$ Corresponding author. E-mail: shanhui@stanford.edu.}
}

\date{}

\begin{document} 


\baselineskip24pt


\maketitle


\begin{sciabstract}
 Convolution is an essential operation in signal and image processing and consumes most of the computing power in convolutional neural networks. Photonic convolution has the promise of addressing computational bottlenecks and outperforming electronic implementations. Performing photonic convolution in the synthetic frequency dimension, which harnesses the dynamics of light in the spectral degrees of freedom for photons, can lead to highly compact devices. Here we experimentally realize convolution operations in the synthetic frequency dimension. Using a modulated ring resonator, we synthesize arbitrary convolution kernels using a pre-determined modulation waveform with high accuracy. We demonstrate the convolution computation between input frequency combs and synthesized kernels. We also introduce the idea of an additive offset to broaden the kinds of kernels that can be implemented experimentally when the modulation strength is limited. Our work demonstrate the use of synthetic frequency dimension to efficiently encode data and implement computation tasks, leading to a compact and scalable photonic computation architecture.
\end{sciabstract}

{{\paragraph*{Teaser:} Light's spectral degrees are used to achieve deterministic and compact photonic convolution in a synthetic frequency dimension.}


\section*{Introduction}\label{sec1}\par Neural networks \cite{LeCun2015} have been ubiquitously employed in machine learning tasks such as computer vision, speech, audio and language comprehension. Among these networks, convolutional neural networks (CNNs) \cite{NIPS2012_c399862d} play a critical role in recognizing features embedded in complex input data. With this capacity of feature extraction, CNNs achieve superior accuracy in predicting unseen data, with a reduced number of parameters compared with dense neural networks \cite{ciregan2012multi}.

\par Convolution, as a central operation for spatio-temporal perception in CNNs \cite{choy20194d}, is particularly energy- and memory-intensive using conventional electronic architecture which is limited by the data movement bottleneck \cite{10.1109/ISCA.2016.40}. As a promising substitute, optical neural networks (ONNs) \cite{Wetzstein2020} process information by propagating a light signal through an optical structure \cite{8859364, Shastri2021}. ONNs have the potential for improved computing performance, with parallel input processing \cite{Feldmann2019}, high computing speed \cite{austrialia_review,Xu2021, Feldmann2021},  broad information bandwidth \cite{Alexoudi2020}, and low energy consumption \cite{Caulfield2010}. ONNs have been implemented in a variety of schemes ranging from Mach–Zehnder interferometers (MZIs) for matrix-vector multiplication \cite{Shen2017, Hughes:18, 8769881} to micro-ring resonators for reservoir computing \cite{Tait2017, PhysRevApplied.11.064043}. Additionally, diffractive layers \cite{Lin1004,Luo2022} and scattering media \cite{Hugheseaay6946} are used for image and vowel detection. However, most of the ONNs are limited to utilizing only the spatial degrees of freedom of photons \cite{Collins2013}. {{Frequency degree of freedom is seldom used for kernel formation \cite{zhao2021scaling}, and most of the previous works utilizing frequency degree of freedom of photons for computing did so without mixing the frequencies as light propagates through the device \cite{Xu2021, Feldmann2021}.}} With the need to scale up ONNs in order to meet the demands of various applications, it is desirable to utilize other degrees of freedom of photons{{, e.g., frequency, }}in order to further enhance the scalability of ONNs'.

\par In this work, we experimentally demonstrate the use of a synthetic frequency dimension \cite{Yuan:18, yuan2021tutorial} as formed by a dynamically modulated ring resonator to enable convolution operations.  Specifically, we synthesize a wide range of convolution kernels with analytically pre-determined modulation waveforms. We achieve various intended convolution kernels with good agreement with theory. We demonstrate the convolution computation by generating different frequency-mode inputs. The output frequency comb obtained from the ring agrees well with the target output as processed by convolution. We also introduce a pathway to broaden the kinds of kernels that can be implemented experimentally when the modulation strength is limited.

The concept of synthetic frequency dimension has been previously employed to demonstrate topological physics \cite{Dutt2019, Dutt59, Hu:20, Wang1240} and matrix-vector multiplication \cite{Buddhiraju2021, zhao2021scaling}. But the use of synthetic frequency dimension for convolution \cite{PhysRevApplied.18.034088} has {{rarely}} been demonstrated experimentally. {{While convolution can be viewed as a special case of matrix-vector multiplication, the matrix that corresponds to a convolution operation has a translational symmetry. Exploiting this translational symmetry enables a simpler implementation as compared with the implementation of a general matrix-vector multiplication operation.} Frequency combs have been previously used for optical convolution purposes \cite{austrialia_review,Xu2021, Feldmann2021}. Refs. \cite{austrialia_review,Xu2021, Feldmann2021} however do not utilize the \textit{dynamics} of light along the frequency dimension, i.e., these works do not utilize the possibility of frequency mixing and conversion as offered by a dynamically modulated system, which is at the heart of the concept of synthetic frequency dimension. Our work introduces a new physics mechanism for achieving optical convolution and is important for the quest to achieve large-scale parallel optical computation with compact devices.

\begin{figure}[ht!]
    \centering
    \includegraphics[width=0.8\textwidth]{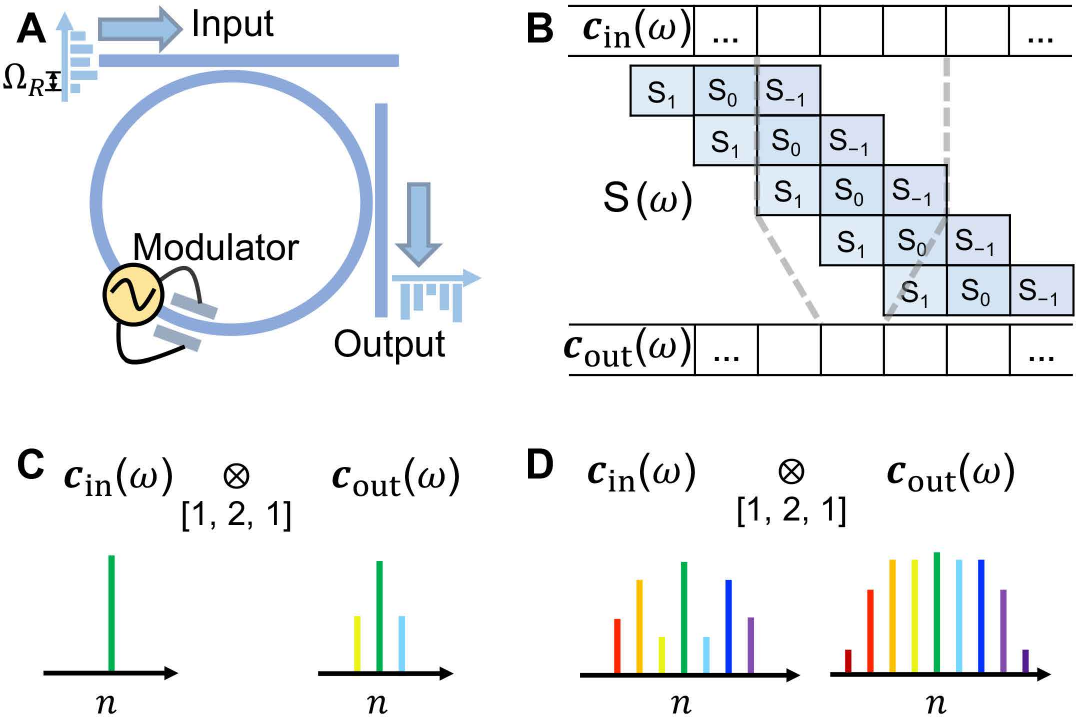}
    \caption{{\bf Schematic illustration of the convolution experiment.} ({\bf A}) The experimental setup, where the convolution operation is performed by a ring resonator modulated by an electro-optical modulator. The modulation has its frequency components located at the free-spectral range $\Omega_R$ of the ring as well as its integer multiples. An input optical frequency comb is injected into the modulated ring resonator. The output frequency comb is detected at the drop-port optical waveguide. ({\bf B}) A translationally symmetric scattering matrix $\VF{S}$ transforms the input $\VF{c}_{\rm in}$ to the output $\VF{c}_{\rm out}$. This transformation is equivalent to a one-dimensional (1D) convolution operation with a kernel. Here we show a three-element kernel $[s_{-1}, s_0, s_{+1}]$ for illustration purposes. ({\bf C}) A continuous-wave laser is injected into a modulated ring resonator implementing a smoothing kernel $[1, 2, 1]$. The output frequency comb manifests the kernel shape. ({\bf D}) A multi-frequency comb is injected into the same modulated ring resonator as in ({\bf C}) that implements the same smoothing kernel $[1, 2, 1]$. The output frequency comb is smoother as compared with the input. In ({\bf C}) and ({\bf D}), the height of a comb line represents the electric field amplitude at the corresponding frequency site.}
    \label{fig:my_label_fig1_expt}
\end{figure}

\section*{Results}

 \paragraph*{Modulation waveform design\label{section_theory}}

\noindent The experimental setup in this work consists of an optical ring resonator undergoing electro-optical modulation as schematically shown in Fig. \ref{fig:my_label_fig1_expt}({\bf A}). Assuming that the waveguide forming the ring resonator, as well as all other waveguides that provide input and output coupling to the ring, all support a single mode and the group velocity dispersion is negligible, $\Omega_R= 2\pi c/n_{\rm g}\ell$ corresponds to the free spectral range (FSR) of the ring resonator. Here $c$, $n_{\rm g}$, and $\ell$ represent light speed in vacuum, group refractive index, and ring circumference, respectively. $t_{R} = 2\pi/\Omega_R$ denotes the round-trip time of the ring. Specifically, here we consider the case that the modulator exclusively modulates the amplitude of light, which can be described by a temporal transmission factor \cite{Wang1240}:
\begin{align}
T_{\rm Am}(t)=\exp\Bigg\{\left[\sum\limits_{m\geq1}B_m\sin(m\Omega_R t+\beta_m) - \gamma t_{R}\right] \Bigg\}
\label{amp_trans_formula}.
\end{align} $B_m$ and  $\beta_m$ corresponds to the magnitude and phase angle of the waveforms in the amplitude modulators for the $m$-th order resonant modulation component, respectively. $\gamma$ corresponds to time-averaged loss as induced by the amplitude modulator. In using Eq. (\ref{amp_trans_formula}) to describe a passive amplitude modulator that has no gain, $\gamma$ is positive and needs to be sufficiently large so that $T_{\rm AM}(t) <1$ for all $t$. The ring resonator is coupled to an input and an output waveguide. Since $T(t) = T(t+t_{R})$, the frequency components of the modulation waveform are located at integer multiples of the FSR of the ring resonator. Therefore, with modulations, the resonant modes of the ring at different frequencies can resonantly couple with each other.

\par In Fig. \ref{fig:my_label_fig1_expt}({\bf A}), there is an input-waveguide that couples to the ring with a coupling coefficient $\gamma_{\rm e1}$, as well as a drop-port waveguide that couples to the ring with a coupling coefficient $\gamma_{\rm e2}$. We inject a frequency comb in the input waveguide to generate an output frequency comb in the drop waveguide. The modulation waveform as described above can be used to implement a convolution kernel in the frequency dimension. The ring resonator supports ${{2N+1}}$ equally spaced resonant modes with frequencies $\omega_n = \omega_0 + n\Omega_R$, $({{-N}}\le n\le N)$, with $\omega_0$ corresponding to the central resonant frequency. We assume an input wave with a form $c_{\rm in} = \sum\limits_n c_{{\rm in},n}\exp{(j \omega_n t + j\Delta \omega t)}$, with $\Delta\omega$ being the detuning. The wave inside the modulated ring then has the form $a(t) = \sum\limits_na_n(t) \exp{(j \omega_n t + j\Delta \omega t)}$. The modal amplitudes $a_n$'s can be determined by the temporal coupled-mode theory \cite{PhysRevApplied.18.034088}. Defining the input and output wave amplitude vectors $\VF{c}_{\rm in} = [\cdots, c_{{\rm in},n}, \cdots]$ and $\VF{c}_{\rm out} = [\cdots, c_{{\rm out},n}, \cdots]$, we obtain the scattering matrix that connects $\VF{c}_{\rm in}$ and $\VF{c}_{\rm out}$ by $\VF{c}_{\rm out} = {\bf S}\VF{c}_{\rm in}$, and ${\bf S}$ is given by,
\begin{equation}
\label{eq:S_expt}
  {\bf S}=j2\gamma_{\rm e1}\gamma_{\rm e2}\{{\bf K}+[ j\left(\gamma+\gamma_{\rm cst} \right)-\Delta\omega]{\bf I}\}^{-1}.
\end{equation}
\noindent Here, $\gamma_{\rm cst}$ is the total rate of loss in the resonator from mechanisms other than the amplitude modulator. These mechanisms can include, for example, the propagation loss of light in the fiber, as well as input and output coupling, as characterized by the input and output coupling rate of $\gamma_{\rm e1}$ and $\gamma_{\rm e2}$, respectively. Here we assume that such a loss rate is the same for every resonant mode in the system. {{This assumption of uniform loss rates holds within a spectral window of a few nanometers, as the variations of coupling ratios for the coupler and the gain profile of amplifiers could be negligible.} $\bf I$ is an identity matrix. The matrix elements of $\bf{K}$ satisfy the translational symmetry, i.e. $K_{m,n} = \kappa_{m-n}$, where $m$ and $n$ are the indices of the modes. $\kappa_{\pm p}$, with $p> 0$, is the coupling constant between two modes $m$ and $n$ satisfying $\lvert m-n\rvert = p$, and is related to the modulation parameters by $\kappa_{\pm p} =  \pm\frac{1}{2t_{R}} B_pe^{\pm j\beta_p}$. To simplify the represention, we denote $\kappa_0 = j\left(\gamma+\gamma_{\rm cst} \right)-\Delta\omega$ to combine the loss and detuning factors into $\VF{K}$ matrix. $\bf{S}$, consequently, is a matrix with elements $S_{m,n} = s_{m-n}$ so it has a translational symmetry along the frequency axis \cite{PhysRevApplied.18.034088}.

\par Due to the translational symmetry along the frequency axis, the scattering matrix in Eq. (\ref{eq:S_expt}) implements a one-dimensional convolution operation,
\begin{align}
    c_{{\rm out},m}&=\sum_{n}S_{m,n}c_{{\rm in},n}=\sum_{n}{s_{m-n}c_{{\rm in},n}}\nonumber\\
    &=\sum_{n}{s_{n}c_{{\rm in},m-n}}.\label{eq_8_conv_first}
\end{align}
Here $s_n$ is the $n$-th element of a kernel for the convolution operation \cite{Jahne2005}. Fig. \ref{fig:my_label_fig1_expt}({\bf B}) illustrates the convolution operation of a kernel consisting of three elements $s_{-1}$, $s_0$, and $s_{+1}$. In this paper, for compactness of notation, we often represent a kernel as a row vector with an odd number of elements. The three-element kernel here for example is denoted as $[s_{-1}, s_0, s_{+1}]$. Fig. \ref{fig:my_label_fig1_expt}({\bf C})-({\bf D}) illustrates the operation of such kernel when $s_0=2$ and $s_{\pm1}=1$. Fig. \ref{fig:my_label_fig1_expt}({\bf C}) corresponds to a case with the input consisting only of a single frequency, namely, the continuous-wave (CW) laser. Fig. \ref{fig:my_label_fig1_expt}({\bf D}) corresponds to a case with the input being a multiple-frequency comb. One-dimensional convolution is essential for feature extraction in sequential data processing such as audio and speech comprehension \cite{KIRANYAZ2021107398}.

\par For experimental design, it would be desirable to generate a prescribed kernel with an analytical modulation waveform. Assuming zero detuning, $\Delta\omega=0$, for a given kernel with elements $s_n$'s, the corresponding modulation parameters $ B_m$ and $\beta_m$ are given by \cite{PhysRevApplied.18.034088},
\begin{align}
B_m\exp({j\beta_m})&=t_{R}(\kappa_{+m}- \kappa^*_{-m}), \label{eq:B_fromKappa_2}\\
\gamma + \gamma_{\rm cst}&= -j\kappa_0,\label{gamma_determine}
\end{align}
\noindent where coupling constants $\kappa_m$ are related to the kernel elements $s_n$'s via \begin{align}\kappa_m =\frac{j\gamma_{\rm e1}\gamma_{\rm e2}}{\pi} \bigintsss_0^{2\pi}\frac{e^{{-jmk}}}{\sum\limits_n e^{jnk}{s_n} }\dif k.\end{align}

\begin{figure*}[ht!]
    \includegraphics[width=\textwidth]{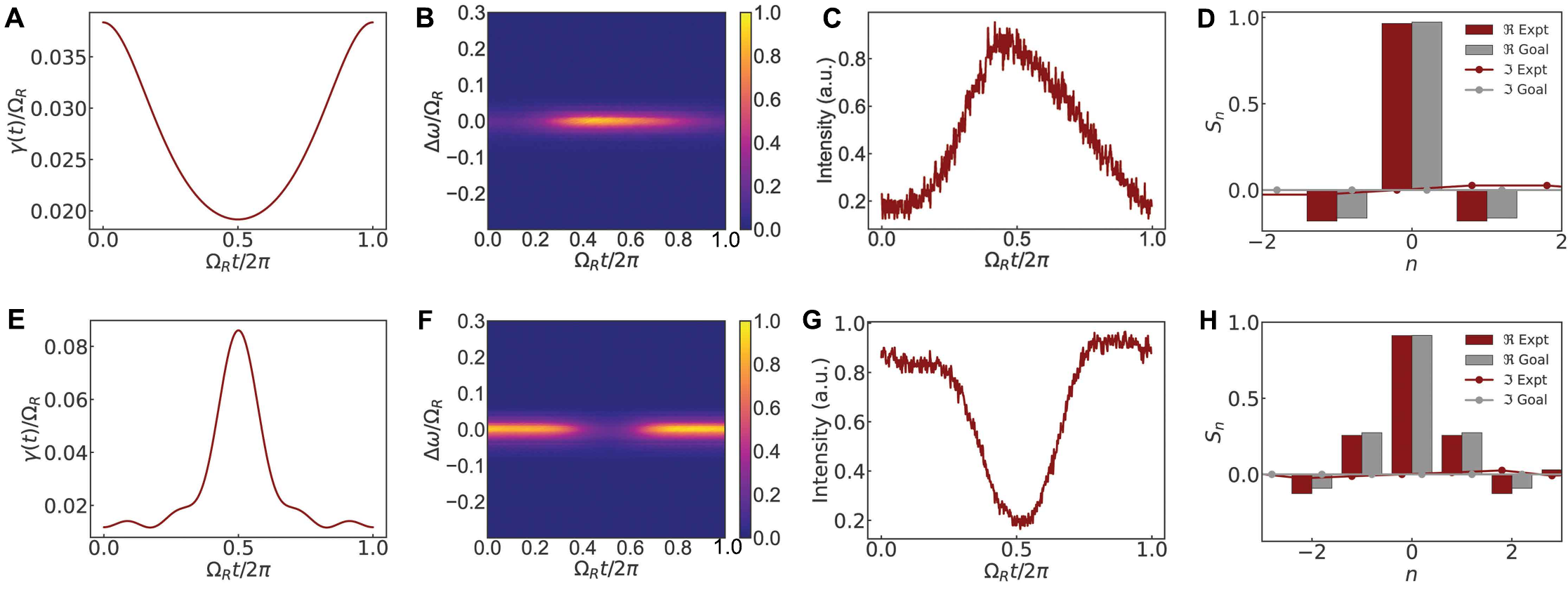}
    \caption{{\bf Experimental synthesis of convolution kernels}. A high-boost kernel $[-1, 6, -1]$ in ({\bf A})-({\bf D}) and a Laplacian of Gaussian kernel $[-1, 3, 10, 3, -1]$ in ({\bf E})-({\bf H}). ({\bf A, E}) Calculated instantaneous loss rate $\gamma(t)$ as a function of time in a roundtrip. ({\bf B, F}) Measured time- and frequency-detuning-resolved output intensity $I(\Delta\omega,t)$. This is measured at the drop port from a dynamically modulated ring resonator. ({\bf C, G}) Measured $I(\Delta\omega = 0,t)$ in (b,f), respectively. ({\bf D, H}) Comparison of the synthesized kernel and target kernel. The red bar/line corresponds to the real/imaginary part of the experimental kernel. The grey bar/line corresponds to the real/imaginary parts of the target kernel.}
    \label{fig:my_label_fig2_expt}
\end{figure*}

\paragraph*{Kernel synthesis\label{sect_synthe}}
\par Our experiments use a fiber ring resonator modulated by an electro-optic modulator as shown in Fig. \ref{fig:my_label_fig1_expt}({\bf A}). The ring has a free spectral range of $\Omega_R = 2\pi\cdot5.99$ MHz, corresponding to a circumference of $\ell = 34.3$ m. From the input waveguide, we launch a continuous wave (CW) laser into the ring resonator through a fiber coupler. The laser's frequency is scanned across a resonance of the unmodulated ring. Within the cavity, we use an Er-doped fiber amplifier (EDFA) to compensate for part of the roundtrip loss. At each detuning $\Delta \omega$, we measure the time-resolved output power $I(\Delta \omega, t)$ at the drop port, using a fast photodiode with a bandwidth over 5 GHz and an oscilloscope of 1 GHz analog bandwidth.

\par We experimentally construct various convolution kernels based on the theory as discussed in the previous Section. Here the {{input-output transformation}} is as described in Fig. \ref{fig:my_label_fig1_expt}({\bf C}) where we launch a single frequency and therefore the output manifests the kernel. In the first example (Fig. \ref{fig:my_label_fig2_expt}({\bf A})-({\bf D})), we demonstrate the high boost kernel, which consists of three nonzero elements of $s_0=6$ and $s_{\pm 1} = -1$. This kernel is widely applied in image processing to sharpen the high-frequency edge information and enhance the low-frequency feature information in the image \cite{Satapathy2019}. 
\par To generate this kernel, we first calibrate the loss rate $\gamma + \gamma_{\rm cst}=0.027\Omega_R$. This calibration is described in more detail in {{Materials and Methods}} Section. With this $\gamma + \gamma_{\rm cst}$ and Eqs. (\ref{eq:B_fromKappa_2})-(\ref{gamma_determine}), we obtain the modulation waveform. For the amplitude modulation as described by Eq. (\ref{amp_trans_formula}), the magnitudes are: $B_1 = $\num{5.858e-2}, $B_2 = $\num{9.994e-3}, $B_3 = $\num{1.679e-3}, $B_4= $\num{2.676e-4}, and $B_5= $\num{3.323e-5}, the phase angles are $\beta_1=1.576$, $\beta_2 = 1.580$, 
$\beta_3=1.585$, $\beta_4=1.590$, $\beta_5 =1.594$. At any given time the instantaneous loss rate of the cavity is defined as \begin{align}
    \gamma(t) =  \gamma+ \gamma_{\rm cst}+\sum\limits_{m\ge1}\frac{B_m}{t_R}\cos(m\Omega_Rt).\label{Modulate_gamma}
\end{align}
\noindent $ \gamma(t)$ is above zero as shown in Fig. \ref{fig:my_label_fig2_expt}({\bf A}). Therefore, the modulation as designed in this way satisfies the passivity constraint \cite{Sandhu_modulation_12} and the system is always dissipative.
 
 \par We apply the modulation waveform, as designed above, to the ring resonator. In the experiment, we vary the detuning $\Delta \omega$ by adjusting the input laser frequency. At each detuning $\Delta \omega$, we record the intensity at the drop port $I(\Delta \omega, t)$ as a function of time. The resulting 2D plot of $I(\Delta \omega, t)$ is plotted in Fig. \ref{fig:my_label_fig2_expt}({\bf B}). We observe that the linewidth of the resonance is the smallest at about $t=\pi/\Omega_R$ in the horizontal axis defined in Fig. \ref{fig:my_label_fig2_expt}({\bf B}). This is consistent with Fig. \ref{fig:my_label_fig2_expt}({\bf A}), where the instantaneous loss rate is lowest at the same $t$. 

\par To determine the kernel from the output intensity measurement $I(\Delta\omega, t)$, we recall that $I(\Delta\omega, t)=\lvert S( \Delta\omega, t)\rvert^2$, with $S(\Delta\omega, t)$ being the time-domain scattering factor of Eq. (\ref{eq:S_expt}). Since the modulation in Fig. \ref{fig:my_label_fig2_expt}({\bf A}) is designed for the kernel at $\Delta\omega = 0$, we plot $I(\Delta\omega=0,t)$ as shown in Fig. \ref{fig:my_label_fig2_expt}({\bf C}). Throughout the paper, all the kernel generation and convolution are based on this $\Delta \omega=0$ line only. As the high-boost kernel demonstrated here is real-valued and symmetric, the time-domain scattering factor $S(0, t)=\sum\limits_ns_n\exp(jn\Omega_Rt)$ should be real-valued as well, with $s_n$ defined in Eq. (\ref{eq_8_conv_first}). In the Supplementary Materials, we use an example single cosine modulation to prove that the modulation waveform only results in a change of $\gamma(t)$ in Eq. (\ref{Modulate_gamma}). We confirm that the amplitude modulation waveform that is obtained from the experiment agrees well with what is implemented on the modulator. This proves that $S(0, t)$ purely results from amplitude modulation, so $S(0, t)$ is real-valued, and the phase variation in a round-trip is negligible. From $I(0, t)$ as shown in Fig. \ref{fig:my_label_fig2_expt}({\bf C}) and $I(0, t)=\lvert S(0, t)\rvert^2$, we obtain $S(0, t)=\sqrt{I(0, t)}$. We then perform a Fourier transform of $S(0, t)$ to determine the kernel $s_n$ that is obtained in the experiment. 

\par In Fig. \ref{fig:my_label_fig2_expt}({\bf D}), we compare the kernel obtained from experiments and target designs. The $s_n$ from the experimental measurement is shown next to the $s_n$ from the target design. Both kernels are normalized such that $\sum\limits_n\lvert s_n\rvert^2=1$. These two kernels agree well and verify that the high-boost kernel is synthesized successfully.

\par As one more example of kernel synthesis, in Figs. \ref{fig:my_label_fig2_expt}({\bf E})-({\bf H}) we synthesize a quantized Laplacian of Gaussian kernel with its non-zero elements being $s_0 = 10$, $s_{\pm1} = 3$, $s_{\pm2} = -1$. This quantized kernel is suitable for compressing features and tracking the machine learning process \cite{NIPS2014_02522a2b}. The modulation waveform is designed in a similar way as above using Eqs. (\ref{eq:B_fromKappa_2})-(\ref{gamma_determine}). The magnitudes of the modulation waveform are $B_1 = 0.1539$, $B_2=0.1014$, $B_3=0.05854$, $B_4 =0.03525$, $B_5=0.02094$. The corresponding phase angles are, $\beta_1 = -1.566$, $\beta_2 = 1.580$, $\beta_3 = -1.557$, $\beta_4 =1.590$, $\beta_5 = -1.547$. Using these parameters, the instantaneous loss rate given by Eq. (\ref{Modulate_gamma}) is plotted in Fig. \ref{fig:my_label_fig2_expt}({\bf E}). Contrary to the prior example, the instantaneous loss rate is highest in the middle of the roundtrip in this case. In Fig. \ref{fig:my_label_fig2_expt}({\bf F}), we present the measured time and frequency detuning resolved output intensity $I(\Delta\omega, t)$. $I(\Delta\omega=0, t)$ is plotted in Fig. \ref{fig:my_label_fig2_expt}({\bf G}). Using a similar method as in the previous example, we extract the experimental kernel $s_n$'s from $I(\Delta\omega=0, t)$. As shown in Fig. \ref{fig:my_label_fig2_expt}({\bf H}), the experimental kernel agrees well with the target kernel, which verifies that our analytically designed modulation waveform can faithfully synthesize a multi-element quantized Laplacian of Gaussian kernel.

\begin{figure*}[ht!]
    \centering
    \includegraphics[width=\textwidth]{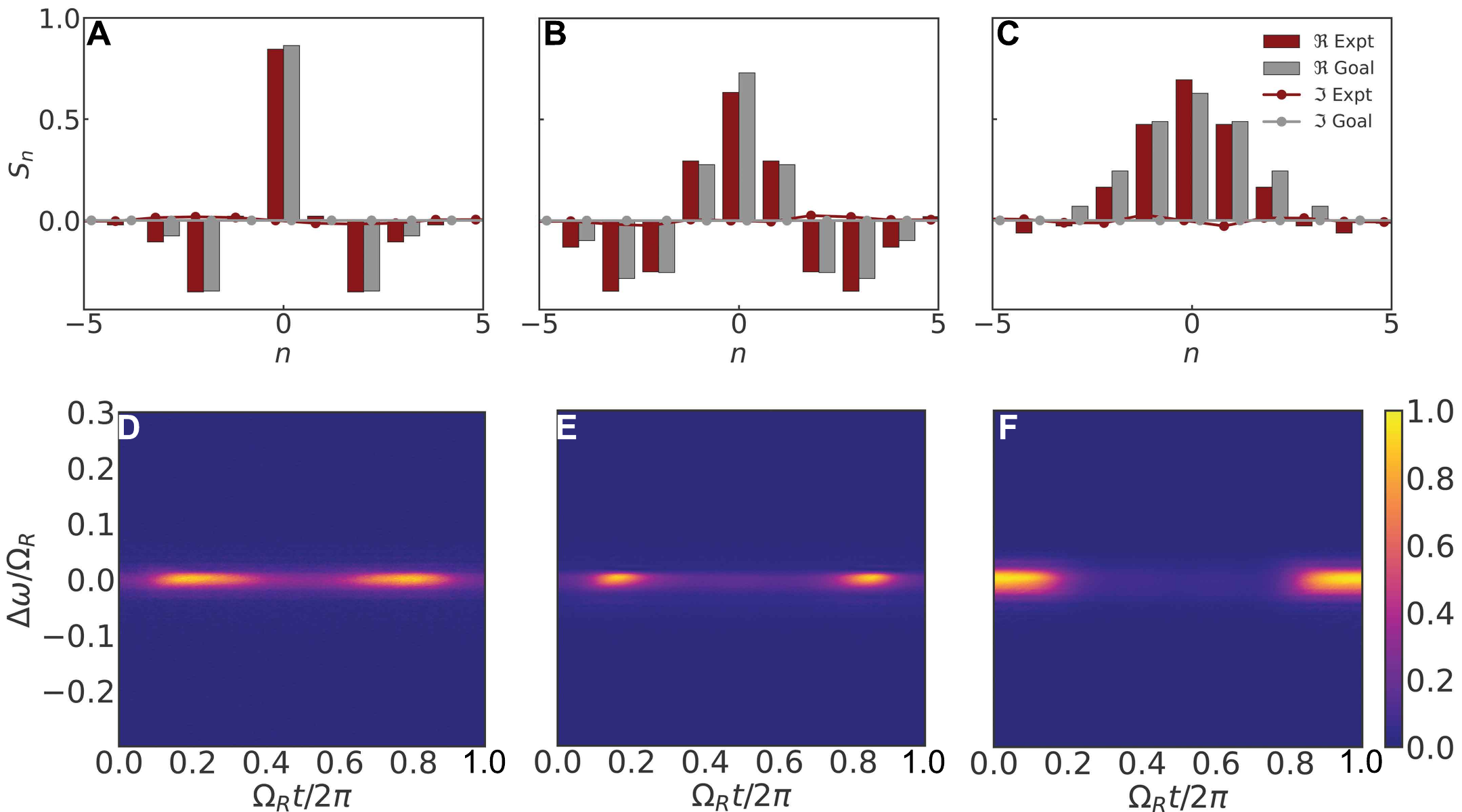}
    \caption{{\bf Construction of convolution kernels with multiple examples of various kernels}. ({\bf A, D}) a standard Laplacian of Gaussian kernel $[-1, -4.56,  0.028, 11.304,  0.028, -4.56, -1]$ with $b=-20$, ({\bf B, E}) another standard Laplacian of Gaussian kernel $[ -1,  -2.9,  -2.6, 2.8, 7.4, 2.8,  -2.6,  -2.9,  -1]$ with $b=-20$, ({\bf C, F}) a Gaussian kernel $[1, 3.5, 7, 9, 7, 3.5, 1]$ with $b=-8$. The upper panels ({\bf A}) to ({\bf C}) correspond to the synthesized kernel measured (in red) and target (in grey) kernels with the real and imaginary parts plotted in bar and lines respectively. The lower panels ({\bf D}) to ({\bf F}) correspond to the time- and frequency-detuning-resolved output intensity measurements. The experimentally synthesized kernel in ({\bf A}) to ({\bf C}) is obtained from ({\bf D}) to ({\bf F}), respectively.}
    \label{fig:my_label_fig3_expt}
\end{figure*}

\paragraph*{Convolution kernel construction with an additive offset}
As seen in the two examples provided in the previous Section, the implemented kernel typically has a strong $s_0$ component in our modulated ring setup. This arises because of the high internal loss factor $\gamma_{\rm cst}$ and the limited modulator strength in our setup. In this Section, we implement the convolution kernel with an additive offset, as described in the form of:
\begin{align}
    c_{{\rm out},n} = b c_{{\rm in},n} + \sum_m s_{n-m} c_{{\rm in},m},\label{bias_term}
\end{align}
where $b<0$ is the additive offset. Alternatively, we consider the implementations of Eq. (\ref{bias_term}) in order to broaden the kinds of kernels that can be implemented in a fiber experimental system. 

\par In our setup, the operation of Eq. (\ref{bias_term}) can be implemented by synthesizing a kernel $\{\tilde{s}_n\}$ where $\tilde{s}_0=s_0 + b$ and $\tilde{s}_n=s_n$ for $n\ne0$, in the same way as we described in the previous Section. On the other hand, in scenarios where the strength of the modulation is insufficeint to directly achieve Eq. (\ref{bias_term}) using the procedure as described in the previous Section, we note that Eq. \ref{bias_term} can be implemented in an alternative all-optical implementation \cite{Minzioni_2019}. In this alternative implementation,  one passes the input light through a beam splitter to separate it into two paths. In the first path, one implements the operation of the first term in Eq. (\ref{bias_term}) using a $\pi$ phase shifter and an attenuator or amplifier. In the second path, one implements the second term in Eq. (\ref{bias_term}) using our modulated fiber ring setup. The transmitted lights from these two paths are then combined to realize Eq. (\ref{bias_term}). A schematic of this realization is provided in the Supplementary Materials. 

\par  Here, as an illustration of Eq. (\ref{bias_term}) and for simplicity, instead of the all-optical implementation as discussed above, we  present results from a hybrid implementation. In the hybrid implementation, for a prescribed target kernel $\{\tilde{s}_n\}$, we separate it into two terms in Eq. (\ref{bias_term}) such that the second term can be implemented using our modulated ring setup. We then present the end results assuming that the first term and the summation operation in Eq. (\ref{bias_term}) have been carried out digitally.

\par In Fig. \ref{fig:my_label_fig3_expt}, we present the implementations of various kernels using this hybrid approach. Both Figs. \ref{fig:my_label_fig3_expt}({\bf A}) and \ref{fig:my_label_fig3_expt}({\bf B}) demonstrate a standard Laplacian of Gaussian kernel with different parameters. In both cases, the kernel elements are summed to zero. This Laplacian of Gaussian kernel is widely applied in noise-robust spatial filtering and edge detection \cite{4767946}. Fig. \ref{fig:my_label_fig3_expt}({\bf A}) corresponds to a seven-element kernel with a standard deviation $\sigma = 1.0$. Fig. \ref{fig:my_label_fig3_expt}({\bf B}) corresponds to a nine-element kernel with a standard deviation $\sigma = 1.4$. Fig. \ref{fig:my_label_fig3_expt}({\bf C}) presents a Gaussian kernel with a standard deviation $\sigma=1.4$. Such a Gaussian kernel is useful for suppressing high-frequency noise in a limited spatial spread area, which is essential for digital telecommunications \cite{1514664}. The details of the modulation waveform parameters can be found in the Supplementary Materials.

\par Figs. \ref{fig:my_label_fig3_expt}({\bf D}) to ({\bf F}) correspond to the time- and frequency-detuning-resolved output intensity measurement. The experimentally synthesized kernel in Figs. \ref{fig:my_label_fig3_expt}({\bf A}) to ({\bf C}) is obtained from Figs. \ref{fig:my_label_fig3_expt}({\bf D}) to ({\bf F}), respectively, using the same method discussed in the previous Section. All of the kernels are normalized such that $\sum\limits_n\lvert s_n\rvert^2 =1$. In Fig. \ref{fig:my_label_fig3_expt}({\bf A}) to Fig. \ref{fig:my_label_fig3_expt}({\bf C}) the measured kernels agree very well with the target kernels in both real and imaginary parts. This verifies that we can synthesize a broad range of kernels at high accuracy with the approach as described by Eq. (\ref{bias_term}). 

\begin{figure*}
    \centering
    \includegraphics[width=\textwidth]{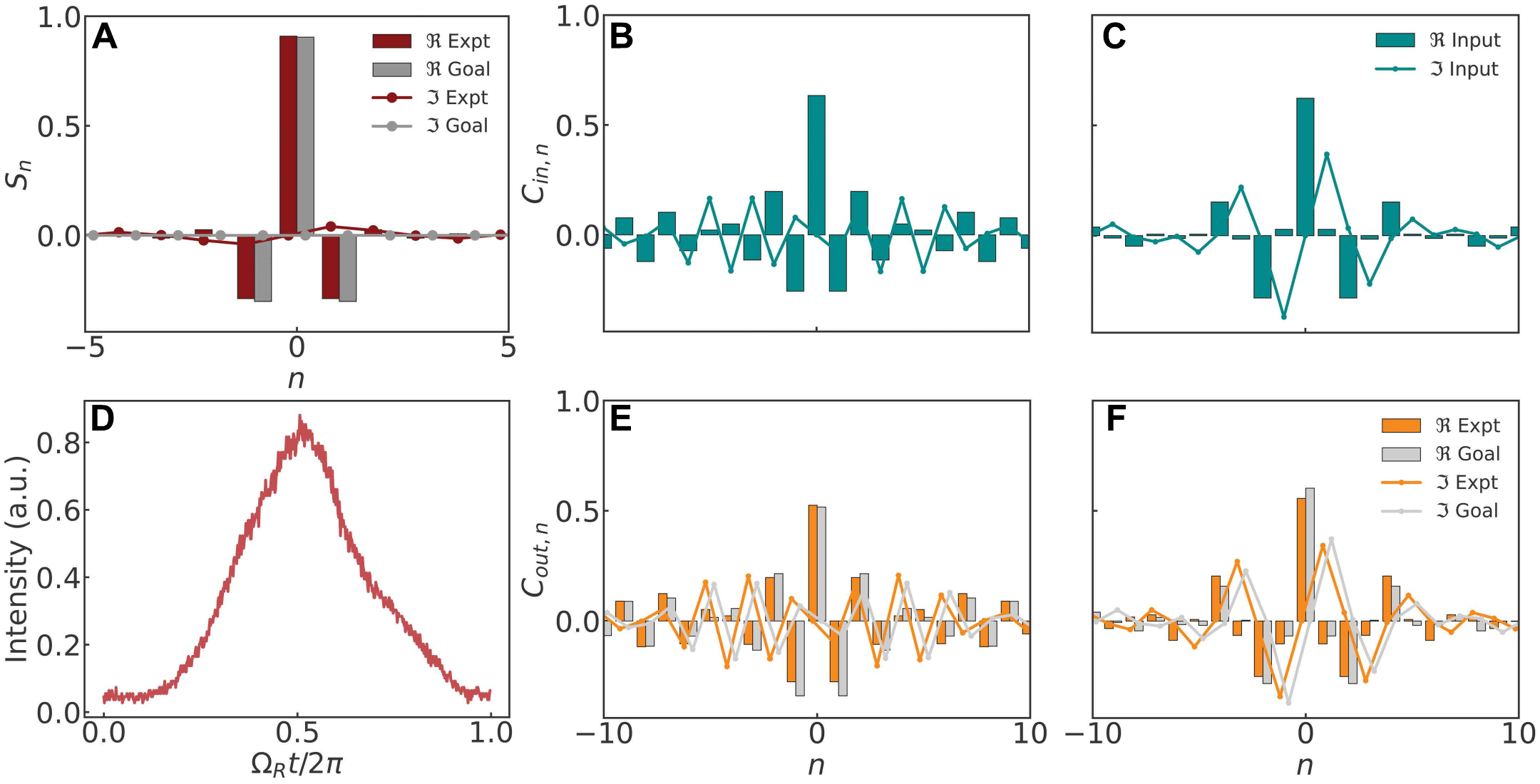}
    \caption{{\bf Convolution processing of the kernels generated from a modulated ring resonator with an input frequency comb consisting of multiple nonzero frequency comb lines.} ({\bf A}) Comparison of the synthesized kernel and target kernel. The red bar/line corresponds to the real/imaginary part of the experimental kernel. The grey bar/line corresponds to the real/imaginary parts of the target kernel. ({\bf B, C}) correspond to the input frequency comb measured from experiments. ({\bf D}) Measured time-resolved intensity from the drop-port of the modulated ring resonator $I(\Delta\omega = 0,t)$ for the kernel synthesis in ({\bf A}). ({\bf E, F}) correspond to the output frequency comb measured (in orange) and expected (in grey) outputs with the real and imaginary parts plotted in bar and lines respectively.}
    \label{fig:my_label_fig4_expt}
\end{figure*}

\paragraph*{Convolution processing}
In the previous sections, we demonstrated the synthesis of several convolution kernels. In these demonstrations, we performed convolution operations with an input vector that consists of only a single element. In this section, we provide an experimental demonstration of the convolution operation of the kernels with various input vectors that consist of multiple frequency comb lines. 

\par To start with we first synthesize a modified Laplacian kernel $s_0 = 3$ and $s_{\pm 1}=-1$. This functions in a similar way as a high boost kernel introduced before, but the reduced $s_0$ term enables an improved edge detection property. We follow the same procedure of applying a pre-determined modulation waveform, as introduced in previous sections.
In Fig. \ref{fig:my_label_fig4_expt}({\bf A}), we compare the kernel obtained from experiments and target designs. The $s_n$ from the experimental measurement is shown next to the $s_n$ from the target design. Both kernels are normalized such that $\sum\limits_n\lvert s_n\rvert^2=1$. These two kernels agree well and verify that the modified Laplacian kernel is synthesized successfully. The slice of $\Delta\omega=0$ in the time- and frequency detuning resolved drop-port intensity measurement is shown in Fig. \ref{fig:my_label_fig4_expt}({\bf D}), which shows a consistent lineshape as in the high-boost kernel case. We emphasize that in this kernel synthesis example, there is no additive offset term involved.

\par To generate the input vector, we use a CW laser operating at a swept frequency across the resonant frequency of the ring and pass the output of the CW laser through an electro-optic amplitude modulator. The modulator is driven by an arbitrary waveform generator (AWG), which has frequency components of the FSR and its integer multiples. This modulation is periodic with a periodicity equal to the round trip time. Such a modulation results in a comb of discrete frequencies equally separated by FSR, which is injected into the ring.

\par The input vector thus generated can be characterized by measuring the time-dependent intensity $I_{\rm in}(t)$ that is transmitted through the modulator. For an amplitude modulator, the amplitude of the transmitted light, up to a global phase that is unimportant, can be determined as $A_{\rm in}(t) = \sqrt{I_{\rm in}(t)}$. A Fourier transform of $A_{\rm in}(t)$ then determines the input vector, i.e. the complex amplitudes of the input light at various frequencies. 

\par Fig. \ref{fig:my_label_fig4_expt}({\bf B}) and  Fig. \ref{fig:my_label_fig4_expt}({\bf C}) show two different input vectors thus generated by applying multiple sinusoidal bands and a sharp pulse, respectively. We choose these two modulations to generate as broadband frequency combs as possible. For each of these input vectors, we send it through the setup corresponding to the kernel shown in Fig. \ref{fig:my_label_fig4_expt}({\bf A, D}). To determine the generated output vector, we measure the output intensity $I_{\rm out}(t)$ as a function of time. Since only the amplitude modulator is used in synthesizing the kernels, we determine the output amplitude $A_{\rm out}(t) = \sqrt{I_{\rm out}(t)}$, we then Fourier transform $A_{\rm out}(t)$ to obtain the output vector. The experimentally determined output vector agrees very well with the direct calculation of the convolution operation of the kernels on the input vectors using the output signal from Fig. \ref{fig:my_label_fig4_expt}({\bf D}), as shown in Fig. \ref{fig:my_label_fig4_expt}({\bf E}) - ({\bf F}). We have thus demonstrated that our setup can indeed achieve convolution operation in the synthetic frequency dimension.

\section*{Discussion}\label{sec12}
\par In summary, we experimentally demonstrate convolution operation in the synthetic frequency space. We show that the prescribed kernel can be implemented by an analytically determined modulation waveform applied to the electro-optic modulator. Our work demonstrates the promise of using frequency to encode data and implement convolution tasks. We anticipate that our demonstration of convolution operation via frequency synthetic dimensions may lead to new types of scalable photonic computation architecture.
 
\par We note that throughout the paper, we only use amplitude modulators, both for the generation of the input signals and for kernel synthesis. As a proof-of-principle experiment, this suffices to demonstrate a wide range of convolution. {{Though we demonstrated some symmetric convolution kernels in our paper, our work can be readily generalized to arbitrary convolution kernels within the capabilities of current experimental setups. For example, an asymmetric kernel can be encoded using an amplitude modulator and a phase modulator in the same ring resonator, as demonstrated in our previous theory work \cite{PhysRevApplied.18.034088}. For kernels with an even number of non-zero elements, we can pad a zero to either side of the kernel and implement it as an asymmetric kernel with an odd number of elements. In addition, to realize kernels with a small $|s_0|$, we discuss in Supplementary Materials an all-optical approach using interference to provide an offset term to the kernel. A deterministic modulation waveform can be obtained in all cases, and we numerically demonstrate two asymmetric kernels in Supplementary Materials. In the case of a general asymmetric kernel, decoding the output signal would require the retrieval of the phase information in the output light, which can be done with the use of heterodyne detection \cite{doi:10.1021/acsphotonics.8b01310}. }}

\par {{The size of the kernel matrix is constrained by the modulation speed bandwidth and FSR, with larger matrices requiring higher modulation speed bandwidth and lower FSR allowing for higher modulation orders. To estimate the performance of our scheme implemented in an on-chip integrated platform, we assume a device with a pump power of 2 mW, 200 input comb lines \cite{Feldmann2021}, modulation speed of 10 GHz, and modulator power cost of 100 mW for a 1 mm$^2$ area chip \cite{Zhang2019}. The computation density for this device is about 4 TOPS mm$^{-2}$, four orders of magnitudes higher than GPU \cite{converged-accelerators}, four times faster than the previous state-of-the-art photonic convolution unit \cite{Feldmann2021}. In terms of power efficiency, we achieve 40 TOPS W$^{-1}$, 6× more power efficiency than Nvidia’s A100X \cite{converged-accelerators}. Our platform is limited only by the photodetector bandwidth. In anticipation, future advances in fabricating high-speed and high-confinement modulators, as well as high-speed photodetectors, will improve our estimations further.
}}
\par {{Our system based on electro-optic modulation complements the previously reported acousto-optic modulation approach \cite{zhao2021scaling}, and can be easily integrated with existing electronic circuitry while allowing a wider range of operating frequencies. All components of the convolution setup demonstrated here can be integrated on a chip. Potential benefits of moving to an integrated platform include lower energy consumption with integrated modulators and lasers, higher computation density with chip-scale compact areas, and more robust and portable edge computing platforms. A major limitation of integration is the large FSR for integrated resonators, which limits the number of modes the modulators and photodetectors can cover at the same time. With the advances in on-chip low-loss waveguides and modulators, it is possible to integrate the entire system without amplifiers and achieve energy-efficient on-chip convolution processes.}}


\section*{Materials and Methods}

\paragraph*{Calibration of the loss rate}\label{secA2_lossfactor}
In this Section, we describe the experimental calibration process of $\gamma + \gamma_{\rm cst}$. Without any modulation from the electro-optical modulator (JDSU model 10020476), we measure the output intensity $I(\Delta\omega)$ from the drop-port of the ring resonator, in the same way as described in Section \ref{sect_synthe}. $I(\Delta\omega)$ is related to $\gamma + \gamma_{\rm cst}$ by, 
\begin{align}
    I(\Delta\omega)=\Big\lvert\frac{2j\gamma_{\rm e1}\gamma_{\rm e2}}{j\gamma + j\gamma_{\rm cst}-\Delta\omega}\Big\rvert^2.
\end{align}
We then perform the least square fitting of $I(\Delta\omega)$ to obtain the optimal parameters of $\gamma + \gamma_{\rm cst}$. In our system, the calibrated loss factor is $\gamma + \gamma_{\rm cst} = 0.027\Omega_R$. We provide more details of extracting the loss factor in the Supplementary Materials.
 
\paragraph*{Data processing and time sequence acquisition}

In our experiments, we use a narrow-linewidth laser with tunable lasing frequency as input (ORION 1550 nm Laser Module) under an amplitude modulator (JDSU, model 10020476) controlled by the radio frequency signal from an Arbitrary Waveform Generator (AWG, AGILENT 33250A-U 80 MHz Function). We use an erbium-doped amplifier (IRE-POLUS, Model EAU-2M) to amplify the optical signal. We use an RF amplifier (Mini-Circuits, Model ZHL-3A+) to amplify the modulation signal. 

\par To measure the time-dependent output intensity $I(\Delta\omega, t)$ at the drop port, we use a photodiode (Thorlabs DET08CFC) with a 5 GHz bandwidth to detect the output signal and we use an oscilloscope (LeCroy LC584AL) with a bandwidth of 1 GHz to obtain a 1-ms time-sequence data. The 1-ms-long time-sequence data was then reshaped into multiple time sequences, one for a roundtrip time of the ring (1/(5.99 MHz) = 167 ns).

\par We determine the starting time of one roundtrip sequence by comparing the intensity peak of the theoretical design peak location. We shift one sequence so that the experimental resonant peak is aligned with the designed peak. The entire measured time sequence is shifted by the same amount of time. We then unflatten the 1D data sequences along the vertical axis to obtain the 2D intensity measurement in Figs. \ref{fig:my_label_fig2_expt}({\bf B, E}) and Figs. \ref{fig:my_label_fig3_expt}({\bf E-H}). The details of the experimental setup can be found in the Supplementary Materials.

\section*{{Acknowledgements}}
We thank Prof. David A.B. Miller for providing the lab space. 

 \paragraph*{Funding}This work is supported by a MURI project from
the U. S. Air Force Office of Scientific Research (AFOSR) (Grant No. FA9550-22-1-0339). 

\paragraph*{Author contributions} Conceptualization: LF, SF.
Methodology: LF, KW, SF.
	Investigation: LF, KW, HW, AD, SF.
	Visualization: LF, SF.
	Supervision: SF.
	Writing—original draft: LF.
	Writing—review $\&$ editing: LF, KW, HW, AD, SF.
 
\paragraph*{Competing interests} LF, KW, SF have filed a patent (US Provisional Patent Application 63/349413 filed 6/6/2023) based on this work. The authors declare no other competing interests.}

\paragraph*{Data and materials availability} All data are available in the main text or the supplementary materials.}

\bibliographystyle{ScienceAdvances}

\end{document}